\documentclass[preprint,sort,compress,12pt]{elsarticle}

\usepackage{amssymb}
\usepackage{amsthm}
\usepackage{amsmath}
\usepackage{mathtools}
\usepackage{mathrsfs}
\usepackage{algorithm}
\usepackage[algo2e]{algorithm2e}
\usepackage{algpseudocode}
\usepackage{array}
\usepackage{multirow}
\usepackage{listings}
\usepackage{tabu}
\usepackage{booktabs}
\usepackage{enumerate}
\usepackage{fullpage}
\usepackage{float}
\usepackage{xcolor}
\usepackage[colorinlistoftodos]{todonotes}
\usepackage{bbm}
\usepackage{bm}
\usepackage[colorlinks=true]{hyperref}
\usepackage{url}
\usepackage{textcomp}
\usepackage{gensymb}
\usepackage{soul}
\usepackage{graphicx}
\usepackage{subfigure}
\usepackage{float}
\usepackage{caption}
\usepackage{tikz}
\usetikzlibrary{shapes}

\theoremstyle{definition}

\theoremstyle{remark}

\biboptions{numbers,comma,round,square}
\graphicspath{ {./figs/} }

\journal{Elsevier}

\begin{document}
\begin{frontmatter}

\title{\emph{Warp-Geo}: Differentiable Geometry Representation for Dynamic-Boundary Simulation and Shape Optimization}



\author[ndAME]{Weijie Zhang\fnref{contrib}} 
\author[cornellMAE]{Xiantao Fan\fnref{contrib}\corref{cor1}} \ead{xf245@cornell.edu} \author[ndAME,cornellMAE]{Jian-Xun Wang\corref{cor2}} \ead{jw2837@cornell.edu} 
\address[ndAME]{Department of Aerospace and Mechanical Engineering, University of Notre Dame, Notre Dame, IN, USA} \address[cornellMAE]{Sibley School of Mechanical and Aerospace Engineering, Cornell University, Ithaca, NY, USA} \fntext[contrib]{Weijie Zhang and Xiantao Fan contributed equally to this work.} \cortext[cor1]{Corresponding author. Tel.: +1 574 413 5927.} \cortext[cor2]{Corresponding author. Tel.: +1 540 315 6512.}

\begin{abstract}
Modern inverse problems and adaptive simulations require efficient and differentiable representations of complex geometries that can be updated frequently for gradient-based optimization. Existing methods cannot simultaneously achieve both. We present \emph{Warp-Geo},  a differentiable GPU-accelerated framework for reconstructing signed distance fields and evaluating surface normals from point clouds of complex 3D geometries. \emph{Warp-Geo} combines uniform grids for GPU parallelism with implicit differentiation through the Poisson solve, enabling frequent geometry recomputation while maintaining full differentiability. We demonstrate \emph{Warp-Geo} on forward SDF reconstruction, dynamic fluid--structure interaction with moving boundaries, and inverse shape optimization via automatic differentiation with gradient validation. GPU scalability analysis confirms practical feasibility for frequent geometry updates. The framework, implemented in \texttt{Warp} and \texttt{JAX}, enables 
seamless integration with differentiable physics solvers for end-to-end shape optimization.
\end{abstract}

\end{frontmatter}

\section{Introduction}
\label{sec:intro}

Shape and geometry representation is fundamental to computational modeling, defining the computational domain and boundary conditions in numerical simulations. However, modern computational workflows increasingly demand that geometry evolves dynamically, at every time step or optimization iteration, while propagating gradients for parameter optimization and adaptive control. This requirement is often overlooked, yet it defines a critical challenge absent from traditional geometry-design pipelines. 

Consider three representative application scenarios. In inverse shape design via differentiable simulation, one optimizes an airfoil or hydrodynamic body to minimize drag or match experimental observations. This requires gradients to propagate backward from the physics solver through geometry to design parameters, necessitating a fully differentiable geometry pipeline where parameters are updated iteratively through gradient-based optimization.
In data assimilation with moving boundaries, the task is to reconstruct cardiovascular geometry from 4D Flow MRI or adapt structure shape to match measured deformations. Geometry must be recomputed and re-evaluated at every assimilation step, making computational efficiency critical.
In adaptive or morphing structures such as swimming fish or morphing aircraft wings, geometry evolves continuously with muscle actuation or flight dynamics. Efficient geometry recomputation at each simulation step is essential; otherwise, geometry operations become the computational bottleneck rather than physics evaluation itself.

The core challenge is enabling frequent geometry updates while propagating gradients. This creates a fundamental tension in existing approaches: approaches optimized for update efficiency sacrifice 
differentiability through non-smooth operations, while approaches that maintain differentiability incur prohibitive computational overhead.
This motivates developing geometry representations that are simultaneously \emph{accurate, efficient to update, GPU-accelerated, and end-to-end differentiable}, enabling high-dimensional geometry optimization required by modern inverse design and adaptive simulation.

Geometry representations can be broadly classified into two categories: explicit and implicit. Explicit representations based on meshes~\cite{sheffer2007mesh, liu2008local, garimella2004triangular}, point clouds~\cite{huang2024surface}, and surface patches~\cite{zhang2025surface} are intuitive and widely supported in CAD pipelines. However, when geometry must be updated dynamically with gradient propagation, explicit methods share a fundamental limitation: they encode geometry through discrete connectivity or sampling patterns that do not differentiate smoothly. Updating an explicit representation requires discrete operations, such as mesh reconnection, point resampling and correspondence, or patch boundary reparameterization, which introduce non-smooth discontinuities that break automatic differentiation. This makes explicit methods poorly suited for high-dimensional geometry optimization where gradients must flow backward through the entire geometric update process.

Implicit representations avoid explicit connectivity by encoding geometry as a continuous scalar field. Common approaches include level-set functions~\cite{osher2003geometric}, signed distance fields (SDFs)~\cite{osher2003signed}, and Poisson-based surface representations~\cite{kazhdan2006poisson,kazhdan2013screened}. Implicit methods offer critical advantages for simulation-oriented applications: no connectivity updates are required, they integrate naturally with fixed-grid solvers and immersed-boundary methods~\cite{peskin2002ibm, mittal2005ibm, shrivastava2013novel} without body-fitted mesh generation, geometric quantities can be evaluated directly from the scalar field via differentiation, and they are particularly useful for moving-boundary simulations such as fluid--structure interaction. However, implicit methods face a critical and often-overlooked challenge when geometry must be updated dynamically and differentiably: the choice between advection-based updating (fast but non-differentiable) versus reconstruction-based updating (differentiable but expensive).

Level-set methods represent moving interfaces as the zero level set of a scalar field $\phi$ using an advection equation $\phi_t + \mathbf{u} \cdot \nabla \phi = 0$, which is computationally efficient for simulating smooth interface motion. However, this advection approach introduces a fundamental barrier to differentiability. After a few advection steps, the solution $\phi$ deviates from a signed distance function; the gradient magnitude $|\nabla \phi|$ no longer equals one everywhere. This deviation causes numerical instability and interface degradation. To maintain accuracy, periodic reinitialization is necessary by solving the Eikonal equation $|\nabla \phi| = 1$  or applying distance-correction procedures. These operations, however, incur significant computational overhead, often dominating the cost of the advection step itself. Moreover, reinitialization involves discrete operations (e.g., sign detection, distance correction) that introduce non-smooth discontinuities incompatible with automatic differentiation.

Poisson surface reconstruction sidesteps the advection-and-reinit problem through a fundamentally different approach: it recomputes the geometry directly from the current point-cloud data, without advection or reinitialization~\cite{kazhdan2006poisson,kazhdan2013screened}. When geometry updates, the reconstruction pipeline is simply run again from scratch. Critically, all operations in this pipeline are smooth, enabling gradients to propagate backward from any downstream objective to the point-cloud input. This makes Poisson-based reconstruction fundamentally suited for end-to-end differentiable geometry optimization.
However, classical Poisson surface reconstruction remains impractical for dynamic scenarios requiring frequent reconstruction. The standard octree-based implementation is geometrically optimal but computationally problematic: octree construction is recursive and irregular, poorly suited to GPU parallelism and automatic differentiation; moreover, the cost of octree reconstruction at each update negates Poisson's efficiency advantage over level-set methods. To realize Poisson's potential, we need a GPU-accelerated, fully differentiable reformulation that maintains geometric fidelity while enabling efficient frequent reconstruction.

Building upon Poisson-based implicit surface reconstruction, we develop \textbf{Warp-Geo}, a differentiable and GPU-accelerated framework for reconstructing signed distance fields from point clouds, specifically optimized for dynamic, frequent geometry updates coupled with gradient-based optimization and inverse design. Compared with traditional CPU-based octree implementations, \emph{Warp-Geo} offers two primary advantages. First, the algorithm is fully vectorized on GPUs, enabling efficient evaluation of SDFs over millions of grid points. The geometry field can be re-evaluated directly from updated point clouds without the repeated advection and reinitialization procedures required by classical level-set methods~\cite{osher2003geometric}. This direct recomputation is fast enough to perform at every time step or optimization iteration, enabling real-time geometry morphing in coupled simulations. Second, the entire geometry-processing pipeline, from point-cloud input through normal estimation, Poisson reconstruction, surface extraction, to SDF computation, is differentiable through automatic differentiation. This allows gradients to propagate consistently from any downstream objective backward to geometry parameters. The framework is naturally compatible with differentiable solvers such as Diff-FlowFSI~\cite{fan2026diff}, DiFVM~\cite{du2026difvm}, and JAX-Fluids~\cite{bezgin2023jax,bezgin2025jax}, and enables end-to-end workflows for simulation, analysis, and gradient-based optimization. Importantly, users can now optimize high-dimensional geometry parameterizations directly, whether sparse control parameters or full point-cloud coordinates, against objectives from coupled physics simulations, without hand-crafting low-dimensional parametrizations or relying on learned surrogates.

Recent work has explored differentiable Poisson reconstruction~\cite{peng2021shape} and learned implicit representations~\cite{park2019deepsdf, ma2020neural}, largely from computer vision perspectives. However, these approaches face critical limitations for dynamic geometry optimization. Learned implicit functions require large training datasets and may not preserve high-fidelity features in complex geometries. Differentiable SDF tools like \texttt{pytorch-volumetric}~\cite{pytorch_volumetric} support field queries but not end-to-end differentiation through reconstruction. For inverse geometry problems requiring frequent reconstruction and high-dimensional gradient propagation, these limitations are prohibitive. \emph{Warp-Geo} addresses these gaps by combining three capabilities: (1) no training data required, guaranteeing high geometric fidelity from first use; (2) GPU-optimized uniform grids enabling frequent recomputation without octree overhead; (3) full reconstruction-process differentiability enabling end-to-end optimization of high-dimensional geometry parameterizations. Unlike level-set 
methods~\cite{osher2003geometric} which require non-smooth reinitialization, \emph{Warp-Geo} maintains differentiability throughout. The framework is constructed entirely from geometric and numerical operations using \texttt{Warp}~\cite{warp2022} and \texttt{JAX}~\cite{jax2018github}.

We demonstrate \emph{Warp-Geo} on multiple representative scenarios. Forward SDF reconstruction from point clouds for geometries spanning diverse domains—fish, aircraft, automobile, jet engine, and cardiovascular geometries—shows reconstruction quality and fidelity to the Eikonal property. Dynamic FSI coupling where geometry is re-evaluated at each time step without advection or reinitialization demonstrates tight integration with differentiable CFD solvers for moving-boundary simulations. Inverse shape optimization via automatic differentiation with comprehensive gradient validation against finite differences shows convergence of parameterized geometry morphing. Computational efficiency analysis demonstrates favorable GPU scalability, confirming practical feasibility for frequent geometry updates.

The remainder of the paper is organized as follows. Section~\ref{sec:methodology} presents the algorithmic pipeline and methodology details. Section~\ref{sec:results} demonstrates forward reconstruction quality on representative geometries, coupling with a differentiable CFD solver for moving-boundary simulations, AD-based shape optimization with gradient validation, and computational cost scaling analysis. Section~\ref{sec:conclusion} summarizes the main findings and outlines future directions, including extensions to adaptive resolution and topology-changing geometries. The source code of \emph{Warp-Geo} is publicly available at \url{https://github.com/jx-wang-s-group/Warp-Geo}.

\section{Methodology}
\label{sec:methodology}

This section details the four modules of \emph{Warp-Geo}: normal-vector estimation, normal-orientation correction, Poisson-system construction, and surface extraction with SDF computation.
\begin{figure}[htp!]
    \centering
   \includegraphics[width=1.0\textwidth]{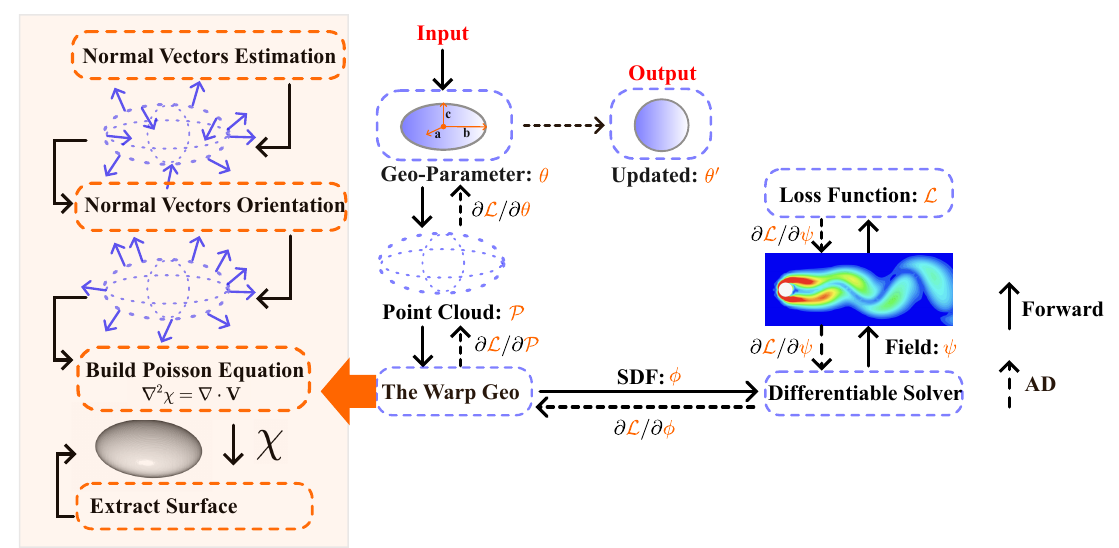}
    \caption{\emph{\emph{Warp-Geo}} for differentiable SDF reconstruction from point clouds, enabling end-to-end integration with differentiable solvers.}
    \label{fig:schematic}
\end{figure}
 As shown in Fig.~\ref{fig:schematic}, the pipeline reconstructs a SDF representation from an unstructured point cloud. Each module is designed for GPU parallelism and automatic differentiation. The algorithmic details and mathematical formulations are presented below.

\subsection{Normal vector estimation}
\label{subsec:normal_estimation}

Poisson surface reconstruction requires an oriented normal vector at each point of the input point cloud. In this work, normal vectors are estimated using distance-weighted principal component analysis (D-PCA), which extracts local tangent-plane information from neighboring point samples. By weighting points inversely by distance, this approach emphasizes nearby samples and suppresses noise at the neighborhood boundary. This procedure is local to each point and is therefore well suited for massively parallel implementation on GPU. 
\begin{figure}[htp!]
    \centering
   \includegraphics[width=0.45\textwidth]{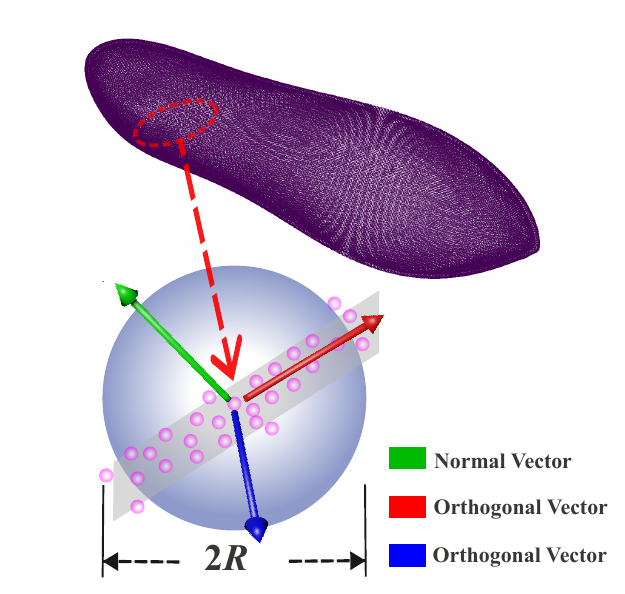}
    \caption{D-PCA based estimation of surface normal vectors from local point-cloud neighborhoods. }
    \label{fig:PCA_vector}
\end{figure}

A schematic illustration of the D-PCA based normal estimation procedure is shown in Fig.~\ref{fig:PCA_vector}. Let the input point cloud be denoted by
$\mathcal{P} = \{\boldsymbol{x}_i\}_{i=1}^{N_p}$, where $\boldsymbol{x}_i$ is the $i^\mathrm{th}$ point-cloud sample and $N_p$ is the total number of points. For each query point $\boldsymbol{x}_i$, we define a local neighborhood $\mathcal{N}_i$ by collecting all points within a sphere of radius $R$:
\begin{equation}
    \mathcal{N}_i =
    \left\{
    \boldsymbol{x}_j \in \mathcal{P}
    \; \big| \;
    \|\boldsymbol{x}_j-\boldsymbol{x}_i\|_2 \leq R
    \right\}.
\end{equation}
Compared with the commonly used $k$-nearest-neighbor search, this radius-based
neighborhood better preserves local geometric symmetry around the query point, e.g., near the tip of a cone. This property is important for complex geometries, where an unbalanced neighbor selection may bias the estimated tangent plane and lead to inaccurate normal directions.

To improve the robustness of the estimation, a distance-weighted covariance matrix is constructed for each query point. Specifically, we define the distance-based weight as
\begin{equation}
    w_{ij} = \frac{1}{\|\boldsymbol{x}_j-\boldsymbol{x}_i\|_2+\epsilon},
\end{equation}
where $\epsilon$ is a small positive constant used to avoid numerical singularity. The weighted centroid of the local neighborhood is then computed as
\begin{equation}
    \bar{\boldsymbol{x}}_i =
    \frac{\sum_{\boldsymbol{x}_j \in \mathcal{N}_i} w_{ij}\boldsymbol{x}_j}
    {\sum_{\boldsymbol{x}_j \in \mathcal{N}_i} w_{ij}} .
\end{equation}
Let $W_i = \sum_{\boldsymbol{x}_j \in \mathcal{N}_i} w_{ij}$. The weighted covariance matrix is then
\begin{equation}
    \boldsymbol{C}_i = \frac{1}{W_i} \sum_{\boldsymbol{x}_j \in \mathcal{N}_i} w_{ij}
    \left(\boldsymbol{x}_j-\bar{\boldsymbol{x}}_i\right)
    \left(\boldsymbol{x}_j-\bar{\boldsymbol{x}}_i\right)^{T}.
    \label{eq:weighted_covariance}
\end{equation}
The distance weighting assigns larger contributions to points closer to the query point and suppresses the influence of distant samples. This reduces over-smoothing of local geometric features and improves the quality of the estimated normals near regions with high curvature or non-uniform point distribution.

The eigendecomposition of $\boldsymbol{C}_i$ gives
\begin{equation}
    \boldsymbol{C}_i \boldsymbol{v}_{i,m}
    =
    \lambda_{i,m} \boldsymbol{v}_{i,m},
    \qquad m=1,2,3,
\end{equation}
where $\lambda_{i,m}$ and $\boldsymbol{v}_{i,m}$ denote the eigenvalues and eigenvectors, respectively. The eigenvectors represent the principal directions of the local point distribution, while the eigenvalues quantify the variance along these directions. For points sampled from a smooth surface, the two eigenvectors associated with the larger eigenvalues span the local tangent plane, whereas the eigenvector associated with the smallest eigenvalue corresponds to the direction of minimum variance. This eigenvector is therefore used as the estimated normal vector,
\begin{equation}
    \boldsymbol{n}_i =
    \boldsymbol{v}_{i,\min},
\end{equation}
where $\boldsymbol{v}_{i,\min}$ is the eigenvector corresponding to the smallest eigenvalue of $\boldsymbol{C}_i$. In practice, the principal directions are computed using singular value decomposition (SVD) of the weighted and centered local point matrix, which is numerically more stable than direct eigendecomposition of the covariance matrix. This stability is critical for automatic differentiation, as direct eigendecomposition becomes ill-conditioned when eigenvalues are nearly equal, producing unstable gradients during backpropagation. SVD avoids this issue and enables reliable gradient propagation through the normal estimation module.

The above procedure is implemented using the single-instruction multiple-thread (SIMT) model enabled by \texttt{Warp}. One GPU thread is assigned to each query point, and all threads independently perform the radius-based neighbor search, weighted covariance construction, eigendecomposition, and normal-vector assignment. Since the computation for each point is independent after the neighbor set is determined, the PCA-based normal estimation can be efficiently parallelized over the entire point cloud.
It should be noted that D-PCA only determines the normal direction up to a sign; namely, both $\boldsymbol{n}_i$ and $-\boldsymbol{n}_i$ are valid solutions of the local decomposition. Therefore, the normals estimated here are unoriented and require a global orientation correction before Poisson surface reconstruction.

\subsection{Parity-based normal orientation correction}
\label{subsec:normal_orientation}
 
Poisson surface reconstruction requires the input normals to be globally consistent, i.e., consistently pointing either outward or inward with respect to the enclosed geometry. Locally flipped normals can lead to holes, discontinuities, or spurious surface patches in the reconstructed surface. In this work, we adopt an outward-pointing convention and correct the D-PCA estimated normals using a parity-based ray-casting procedure. 

\begin{figure}[hpt!] 
\centering 
\includegraphics[width=0.5\textwidth]{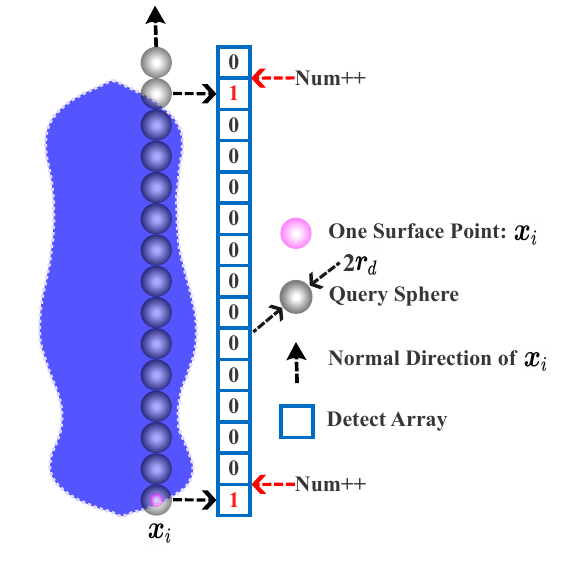} 
\caption{Parity-based correction of normal-vector orientation. A ray is cast along the D-PCA estimated normal direction, and the parity of the detected surface crossings is used to determine the correction.} 
\label{fig:normal_adjust} 
\end{figure}

\paragraph{Basic parity-based orientation test}
The correction is based on the standard inside--outside test for closed geometries, commonly implemented through the ray-casting or crossing-number criterion~\citep{shimrat1962algorithm,berg2000computational}. According to this criterion, a ray starting from an exterior point intersects the boundary an even number of times, whereas a ray starting from an interior point gives an odd number of intersections. Therefore, the sign of a D-PCA estimated normal can be determined by testing whether a ray cast along the normal direction starts from the exterior or interior side of the surface. As illustrated in Fig.~\ref{fig:normal_adjust}, for each point $\boldsymbol{x}_i$ with its D-PCA estimated normal $\boldsymbol{n}_i$, we cast a ray from the point along the normal direction: 
\begin{equation} 
\boldsymbol{r}_i(s) = 
\boldsymbol{x}_i + s\boldsymbol{n}_i, \qquad s \geq 0 . \end{equation} 
If $\boldsymbol{n}_i$ points outward, the ray leaves the geometry immediately after the initial surface point. If $\boldsymbol{n}_i$ points inward, the ray first enters the geometry and must cross the surface again before reaching the exterior. Thus, after excluding the initial surface intersection associated with $\boldsymbol{x}_i$, an even number of remaining intersections indicates an outward normal, whereas an odd number indicates an inward normal. In the latter case, the normal vector is flipped. 

\paragraph{Point-cloud ray casting with finite-radius detection}
Unlike mesh-based ray casting, the present geometry is represented only by a discrete point cloud. Therefore, ray--surface intersections cannot be computed exactly from triangle intersections. Instead, we detect them approximately through local point-cloud occupancy. Along each ray, a sequence of probe points is sampled with a uniform spacing $\Delta s$: 
\begin{equation} 
\boldsymbol{y}_{i,m} = \boldsymbol{x}_i + m\Delta s \boldsymbol{n}_i, \qquad m=0,1,\ldots,M . 
\end{equation} 
For each probe point, we check whether there exists at least one point-cloud sample within a detection radius $r_d$. This defines a binary detection array, 
\begin{equation} 
d_{i,m} = \begin{cases} 
1, 
& 
\exists \boldsymbol{x}_j \in \mathcal{P} \ \text{such that}\ \|\boldsymbol{x}_j-\boldsymbol{y}_{i,m}\|_2 \leq r_d, \\ 0, & \text{otherwise}. 
\end{cases} 
\end{equation} 
A continuous block of nonzero entries in this array corresponds to one detected surface-crossing region. Therefore, the number of detected crossings can be counted by identifying transitions from $1$ to $0$ in the detection array: 
\begin{equation} 
N_i^{\mathrm{all}} = \sum_{m=1}^{M} \mathbb{I} \left( d_{i,m}=0 \ \text{and}\ d_{i,m-1}=1 \right), 
\end{equation} 
where $\mathbb{I}(\cdot)$ is the indicator function. Since the first detected block corresponds to the original surface point $\boldsymbol{x}_i$, it is excluded from the parity check. The effective crossing number is therefore defined as 
\begin{equation} 
N_i = N_i^{\mathrm{all}} - 1 . 
\label{eq:N_i_define} 
\end{equation} 
The basic orientation correction is then 
\begin{equation}
    \boldsymbol{n}_i \leftarrow
    \begin{cases}
        \phantom{-}\boldsymbol{n}_i,  & N_i \bmod 2 = 0, \\
        -\boldsymbol{n}_i, & N_i \bmod 2 = 1.
    \end{cases}
    \label{eq:basic_parity_flip}
\end{equation}
Here, the first case indicates that the D-PCA estimated normal is already outward-facing, while the second case indicates an inward-facing normal that must be flipped. The detection radius is initialized from the point-cloud spacing: 
\begin{equation} r_{d,0} = \frac{1}{2} \max_i \left( \min_{j\neq i} \|\boldsymbol{x}_i-\boldsymbol{x}_j\|_2 \right). 
\label{eq:baseline_detection_radius} 
\end{equation} 
This value provides a conservative estimate of the largest local point spacing. In practice, we introduce an expansion factor $\gamma_r$ and set 
\begin{equation} 
r_d = \gamma_r r_{d,0}, \qquad \gamma_r \geq 1 , 
\label{eq:expanded_detection_radius} 
\end{equation} 
which improves the robustness of crossing detection for sparse point clouds. 

\begin{figure}[t!] 
\centering 
\includegraphics[width=1.0\textwidth]{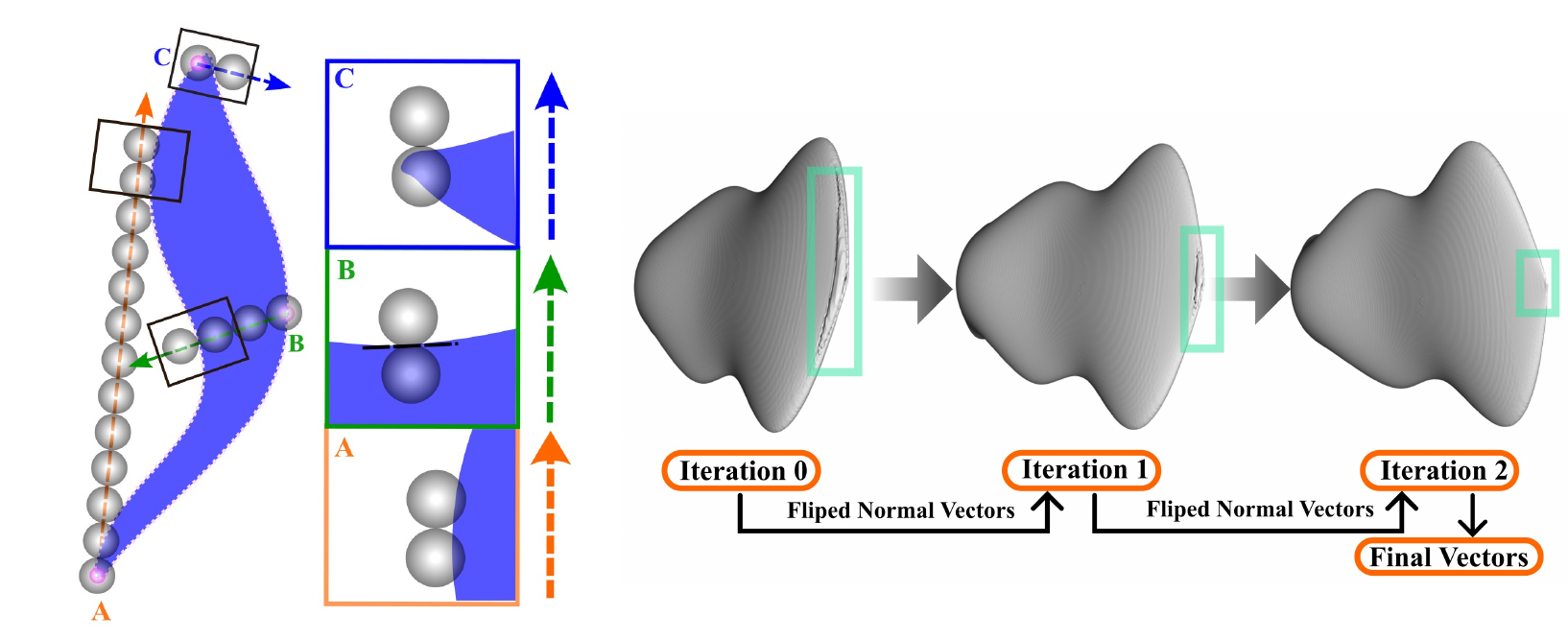} 
\caption{Iterative ray-casting refinement for normal-orientation correction near sharp and thin regions. (a) Typical failure modes caused by finite-radius point-cloud detection. (b) Improvement of local normal consistency using the proposed iterative crossing-number criteria based on $N_i$.} 
\label{fig:sharp_refined} 
\end{figure}

\paragraph{Iterative refinement for sharp and thin features.}
The main practical difficulty of the above point-cloud ray-casting procedure is the finite detection radius. As shown in Fig.~\ref{fig:sharp_refined}, three typical failure modes may occur. In scenario~A, a large detection radius produces a false crossing because the query sphere detects nearby points even though the ideal zero-width ray does not intersect the surface. In scenario~B, the ray intersects the surface, but the local point distribution is nearly tangent to the detection spheres, causing the crossing to be missed. In scenario~C, two nearby surface layers in a sharp or thin region are merged into a single detected block, leading to an incorrect crossing number. Among these cases, scenario~B occurs rarely and can generally be avoided by adjusting the sphere radius. To address scenarios~A and C, we introduce an iterative ray-casting refinement based on the effective crossing number $N_i$. This refinement is one of the key algorithmic contributions of \textit{Warp-Geo}. Instead of relying on a single parity check, we repeatedly evaluate the detection pattern and apply additional criteria to correct ambiguous normals near sharp surfaces and thin regions. The normal vector is updated iteratively as follows:
\begin{itemize}
    \item \textit{Iteration 0 (Standard parity).} Apply the standard parity rule: flip if $N_i \bmod 2 = 1$.
    \item \textit{Iteration 1 (Ambiguous detection pattern).} For normals that may have ambiguous detection patterns (e.g., very long initial nonzero block), apply additional criterion: flip if $N_i \neq 0$ or $N_{i,\mathrm{one}} > N_{\mathrm{tol}}$, where $N_{i,\mathrm{one}}$ denotes the number of consecutive nonzero entries at the beginning of the detection array, and the threshold $N_{\mathrm{tol}}$ is a user-defined tolerance, set to $2$ by default.
    \item \textit{Iteration 2 (Symmetric comparison).} Compute the crossing number $N_i^{-}$ in the opposite direction and select the direction with fewer crossings: flip if $N_i > N_i^{-}$.
\end{itemize}
These iterations are applied in sequence, with each refining the normal orientation based on increasingly detailed inspection of the detection pattern.

The entire orientation-correction procedure is implemented in the SIMT model using \texttt{Warp}. One GPU thread is assigned to each point-cloud sample, and each thread independently performs ray sampling, local occupancy detection, crossing-number evaluation, and normal flipping. The method is therefore fully parallel over the point cloud and does not require explicit surface connectivity or a precomputed mesh. The corrected normals are then used in the subsequent Poisson reconstruction step.

The main user-adjustable hyperparameters of \emph{Warp-Geo} are summarized in Table~\ref{tab:hyperparameters}. These parameters control the locality of normal estimation, the robustness of ray-casting-based orientation correction, and the resolution of the background Eulerian grid. The default values provide a practical starting point, and they can be adjusted for geometries with different point-cloud densities, sharp features, or thin structures.
\begin{table}[H]
\centering
\caption{Main user-adjustable hyperparameters in \emph{Warp-Geo}.}
\label{tab:hyperparameters}
\begin{tabular}{lll}
\hline
Hyperparameter & Symbol & Default \\
\hline
D-PCA neighborhood radius & $R$ & 0.1 \\

Ray-casting detection radius & $r_d$ & Eq.~\ref{eq:baseline_detection_radius} \\

Detection-radius expansion factor & $\gamma_r$ & 2.5 \\

Grid density & $d$ & 8.0 \\

Initial-one threshold & $N_{\mathrm{tol}}$ & $2$ \\
\hline
\end{tabular}
\end{table}

\subsection{GPU-efficient Poisson reconstruction on uniform grids}
\label{subsec:poisson_equation}

After normal-vector orientation correction, the geometry reconstruction is formulated as a Poisson problem. The objective is to recover an implicit indicator function $\chi$ whose gradient aligns with the oriented normal vector field. The governing equation is
\begin{equation}
    \nabla^2 \chi = \nabla \cdot \boldsymbol{V},
    \label{eq:poisson}
\end{equation}
where $\boldsymbol{V}$ is a vector field constructed from the outward-pointing normals $\{\boldsymbol{n}_i\}$ estimated in the previous sections. Solving this equation yields an implicit representation of the surface as a level set of $\chi$.

\paragraph{Galerkin discretization on uniform Eulerian grid}
We discretize the domain $\Omega$ (the bounding box of the point cloud) using a uniform Eulerian grid with $l \times m \times n$ cells of uniform size $h$. The implicit function is approximated by a Galerkin expansion:
\begin{equation}
    \chi_h(\boldsymbol{q}) = \sum_{o \in \mathcal{G}} a_o F_o(\boldsymbol{q}),
    \label{eq:chi_expansion}
\end{equation}
where $\boldsymbol{q}$ denotes a spatial query location, $\mathcal{G}$ is the set of grid cells, $F_o(\boldsymbol{q})$ is the basis function associated with cell $o$ (typically tri-linear), and $a_o$ are unknown coefficients to be solved. The basis function is defined as
\begin{equation}
    F_o(\boldsymbol{q}) = \frac{1}{h^3} F\left(\frac{\boldsymbol{q}-\boldsymbol{c}_o}{h}\right),
    \label{eq:basis_function}
\end{equation}
where $\boldsymbol{c}_o$ is the cell center and $F$ is a compactly supported kernel. The compact support ensures that $F_o$ is nonzero only within a $3 \times 3 \times 3$ neighborhood of cell $o$, which is critical for computational efficiency.

\paragraph{Weak formulation and linear system}
Instead of directly discretizing Eq.~\eqref{eq:poisson}, we use the equivalent variational formulation. The reconstruction minimizes the energy
\begin{equation}
    E(\chi_h) = \frac{1}{2} \int_{\Omega} \left\| \nabla \chi_h(\boldsymbol{q}) - \boldsymbol{V}(\boldsymbol{q}) \right\|^2 \, d\boldsymbol{q},
    \label{eq:poisson_energy}
\end{equation}
which measures how well the gradient of $\chi_h$ matches the normal field $\boldsymbol{V}$. Taking the variation with respect to each basis function $F_p$ (as test function in the Galerkin method) yields
\begin{equation}
    \int_{\Omega} \nabla F_p(\boldsymbol{q}) \cdot \left[\nabla \chi_h(\boldsymbol{q}) - 
    \boldsymbol{V}(\boldsymbol{q})\right] \, d\boldsymbol{q} = 0.
    \label{eq:weak_variation}
\end{equation}
This weak form is equivalent to the Poisson equation with homogeneous Neumann boundary conditions (zero normal flux at domain boundaries). Substituting the Galerkin expansion yields the linear system
\begin{equation}
    \boldsymbol{A}\boldsymbol{a} = \boldsymbol{b},
    \label{eq:linear_system}
\end{equation}
where $\boldsymbol{a}$ is the vector of unknown coefficients. and the entries of the stiffness matrix $\boldsymbol{A}$ are given by
\begin{equation}
    A_{p,o}
    =
    \left\langle
    \nabla F_p, \nabla F_o
    \right\rangle
    =
    \int_{\Omega}
    \nabla F_p(\boldsymbol{q})
    \cdot
    \nabla F_o(\boldsymbol{q})
    \, d\boldsymbol{q},
    \label{eq:lhs_matrix}
\end{equation}
which depends only on the grid structure and basis functions, independent of the point cloud. The right-hand-side vector is
\begin{equation}
    b_p
    =
    \left\langle
    \nabla F_p, \boldsymbol{V}
    \right\rangle
    =
    \int_{\Omega}
    \nabla F_p(\boldsymbol{q})
    \cdot
    \boldsymbol{V}(\boldsymbol{q})
    \, d\boldsymbol{q}.
    \label{eq:rhs_vector}
\end{equation}
which is directly determined by the distributed point normals (detailed below).

\begin{figure}[htp!]
    \centering
    \includegraphics[width=1.0\textwidth]{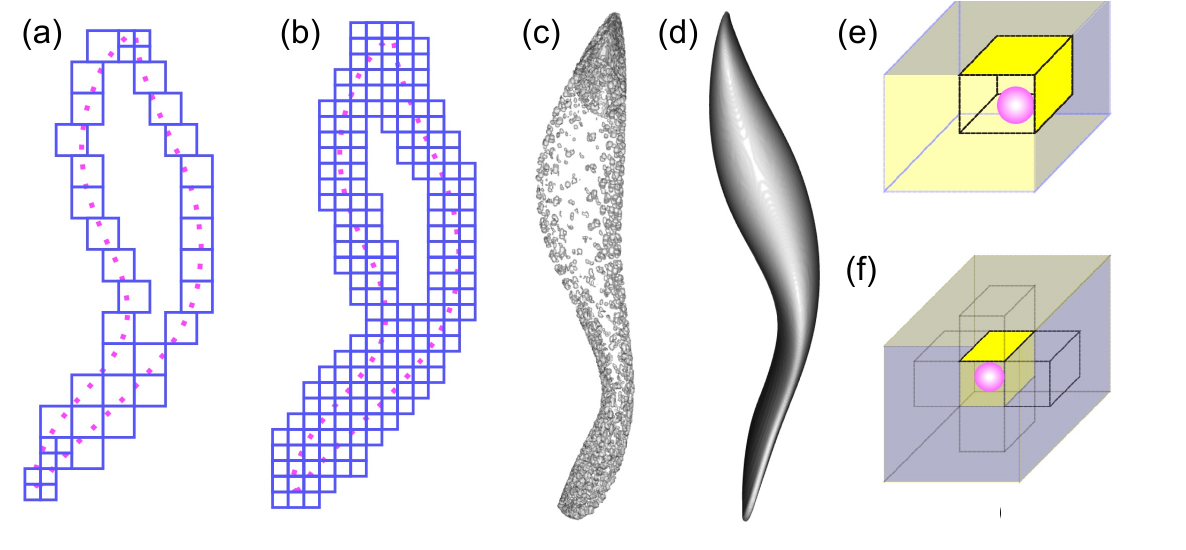}
    \caption{Comparison of cell-selection strategies for Poisson surface reconstruction. 
    (a) Sparse octree-style cell selection using a $2 \times 2 \times 2$ stencil.
    (b) Uniform-grid cell selection with a $3 \times 3 \times 3$ stencil.
    (c) Imperfect surface from sparse $2 \times 2 \times 2$ uniform-grid stencil.
    (d) Smooth surface from proposed $3 \times 3 \times 3$ stencil.
    (e) Naive uniform-grid formulation: each point normal distributed to only the containing cell.
    (f) Proposed approach: each point normal distributed to surrounding $3 \times 3 \times 3$ cells.} 
    \label{fig:grid_selection}
\end{figure}

\paragraph{GPU advantages of uniform grid discretization}
Classical Poisson surface reconstruction uses adaptive octrees for geometric adaptivity (Fig.~\ref{fig:grid_selection}(a)): finer cells near features, coarser cells in empty regions. It is effective for adaptively resolving regions with different point densities and provides a hierarchical basis for the Galerkin discretization. However, octrees present significant obstacles to GPU parallelization. The recursive tree construction introduces dynamic control flow; cells at different tree depths have variable sizes, leading to irregular memory access patterns; and workload is non-uniform across GPU threads. More critically, the discrete tree structure 
and adaptive refinement introduce conditional branching and scale-dependent operations that complicate gradient propagation through AD, making it difficult to maintain smooth gradient flow for end-to-end optimization.

A uniform grid eliminates these obstacles. All cells are identical and regularly arranged in memory, enabling predictable memory access patterns and balanced thread workloads. For a static geometry, the primary efficiency advantage arises from the structure of the linear system: the matrix $\boldsymbol{A}$ can be assembled and factorized once and then reused, while only the right-hand-side vector $\boldsymbol{b}$ is updated when appropriate. For moving or deforming geometries, however, the basis functions in Eq.~\ref{eq:basis_function} change with the point cloud, requiring $\boldsymbol{A}$ to be reassembled. Nevertheless, the topology of the computational graph remains fixed, allowing gradients to propagate consistently through the geometry-dependent matrix assembly and linear solve. Moreover, the regular grid avoids the discrete branching associated with adaptive methods, thereby facilitating reliable gradient-based shape optimization and moving-boundary simulations.

\paragraph{Right-hand side: robust normal distribution with 27-cell stencil}
The geometric information from the point cloud enters the linear system through the right-hand-side vector (Eq.~\eqref{eq:rhs_vector}). For each oriented point $(\boldsymbol{x}_i, \boldsymbol{n}_i)$, the normal is distributed to neighboring grid cells:
\begin{equation}
    \boldsymbol{V}(\boldsymbol{q}) = \sum_{i=1}^{N_p} \sum_{o \in \mathcal{C}_i}
    \alpha_{o,i} F_o(\boldsymbol{q}) \boldsymbol{n}_i,
    \label{eq:V_field}
\end{equation}
where $\mathcal{C}_i$ is the set of neighboring cells and $\alpha_{o,i}$ are interpolation weights.
A naive implementation distributes each normal to only its containing cell, as illustrated in Fig.~\ref{fig:grid_selection}(e). However, for sparse or non-uniformly distributed point clouds, this produces weak local coupling: nearby surface samples activate basis functions with little overlap. This weak coupling can produce fragmented implicit fields and incomplete reconstructed surfaces (see Fig.~\ref{fig:grid_selection}(c)).

To overcome this limitation while maintaining GPU efficiency, we introduce a dense local interpolation strategy on the uniform Eulerian grid. For each point, $\mathcal{C}_i$ is defined as the $3 \times 3 \times 3$ stencil of 27 cells surrounding the point, as illustrated in Figs.~\ref{fig:grid_selection}(b) and~\ref{fig:grid_selection}(f). The normal $\boldsymbol{n}_i$ is distributed to all 27 cells using normalized tri-linear interpolation weights:
\begin{equation}
    \alpha_{o,i} = \frac{\tilde{\alpha}_{o,i}}{\sum_{o' \in \mathcal{C}_i} \tilde{\alpha}_{o',i}},
    \label{eq:alpha_normalized}
\end{equation}
where the unnormalized weight for each cell is
\begin{equation}
\tilde{\alpha}_{o,i}=
\begin{cases}
  \dfrac{1}{2},
  & \text{if } o \text{ is the central cell in } \mathcal{C}_i, \\[3mm]
  \displaystyle
  \prod_{\xi \in \{x,y,z\}}
    \frac{1}{2}
    \max
    \left(
    0,
    1 -
    \frac{|x_{i,\xi}-c_{o,\xi}|}{2h}
    \right),
  & \text{otherwise.}
\end{cases}
\label{eq:alpha_product}
\end{equation}
Here, $x_{i,\xi}$ and $c_{o,\xi}$ denote the $\xi$-direction coordinates of point $\boldsymbol{x}_i$ and cell center $\boldsymbol{c}_o$, respectively. Cells closer to the point receive larger weights, forming a tri-linear interpolation pattern. This dense stencil enlarges the effective support of each point normal, strengthening the local coupling between basis functions. As shown in Fig.~\ref{fig:grid_selection}(d), the result is smooth, complete surfaces even for sparse point clouds. Importantly, the stencil size is fixed (27 cells regardless of grid resolution), so memory layout remains regular and GPU-friendly. This design combines the robustness advantages of classical Poisson methods with the parallelization efficiency of modern GPUs.
Substituting the distributed normal field (Eq.~\eqref{eq:V_field}) into the right-hand-side vector (Eg.~\eqref{eq:rhs_vector}) yields
\begin{equation}
    b_p
    =
    \sum_{i=1}^{N_p}
    \sum_{o \in \mathcal{C}_i}
    \alpha_{o,i}
    \int_{\Omega}
    \nabla F_p(\boldsymbol{q})
    \cdot
    \left[
    F_o(\boldsymbol{q})\boldsymbol{n}_i
    \right]
    \, d\boldsymbol{q}.
    \label{eq:rhs_discrete}
\end{equation}
After solving the linear system (Eq.~\eqref{eq:linear_system}), the coefficients $\boldsymbol{a}$ define the implicit field $\chi_h$, which is subsequently used for surface extraction and SDF computation.

\subsection{Surface extraction and signed-distance-field computation}
\label{subsec:surface_sdf}
Solving the Poisson system in Eq.~\eqref{eq:linear_system} gives the scalar implicit field $\chi_h$ over the computational domain. This field provides the reconstructed geometry through one of its isosurfaces,
\begin{equation}
    \Gamma_{\chi}
    =
    \left\{
    \boldsymbol{q} \in \Omega
    \; \big| \;
    \chi_h(\boldsymbol{q}) = c
    \right\},
    \label{eq:isosurface}
\end{equation}
where $c$ is the selected iso-value. Following the standard Poisson surface reconstruction procedure, $c$ is determined from the reconstructed indicator values near the input point samples~\citep{kazhdan2006poisson}. The extracted surface $\Gamma_{\chi}$ is then used as the geometric interface from which the final signed distance field (SDF) is computed. The SDF is a distance-based 
reinitialization of the Poisson implicit field that preserves the reconstructed iso-surface while replacing implicit indicator values with signed Euclidean distances.

\paragraph{Differentiable isosurface extraction}
The isosurface $\Gamma_{\chi}$ is extracted from $\chi_h$ using the Marching Cubes algorithm~\citep{lorensen1987marching}. Marching Cubes operates on the regular grid where $\chi_h$ is defined. Each grid cell contains eight vertices, and the scalar value of $\chi_h$ at these vertices determines whether the iso-value $c$ is crossed inside the cell. If an edge has one endpoint with $\chi_h>c$ and the other with $\chi_h<c$, the surface intersects that edge. The intersection point is obtained by linear interpolation,
\begin{equation}
    \boldsymbol{q}_{e}
    =
    (1-\eta)\boldsymbol{q}_1 + \eta \boldsymbol{q}_2,
    \qquad
    \eta =
    \frac{c-\chi_h(\boldsymbol{q}_1)}
    {\chi_h(\boldsymbol{q}_2)-\chi_h(\boldsymbol{q}_1)} ,
    \label{eq:marching_cube_interp}
\end{equation}
where $\boldsymbol{q}_1$ and $\boldsymbol{q}_2$ are the two endpoints of the crossing edge. The local triangle connectivity is determined from the standard Marching Cubes lookup table according to the inside--outside configuration of the eight cell vertices.

The Marching Cubes procedure is provided by \texttt{Warp}. Since each grid cell can be processed independently, the algorithm is naturally parallel over the computational grid. For a fixed cell configuration, the interpolated vertex locations in Eq.~\eqref{eq:marching_cube_interp} are differentiable with respect to the scalar field values of $\chi_h$. Therefore, the surface extraction step is piecewise differentiable and can be integrated into the proposed differentiable geometry-processing pipeline. The non-smoothness arises only when the discrete Marching Cubes case changes, corresponding to a change in the local iso-surface topology.

\paragraph{GPU-parallel signed distance reinitialization}

Although $\chi_h$ is a volumetric scalar field, it is not itself a signed distance field. Its values lack geometric meaning, and it does not satisfy the Eikonal property required by downstream solvers. Therefore, after extracting the surface $\Gamma_{\chi}$, we convert $\chi_h$ into a proper signed distance field $\phi$ whose zero level set coincides with $\Gamma_{\chi}$ while providing metrically meaningful distance information:
\begin{equation}
    \phi(\boldsymbol{q})
    =
    \sigma(\boldsymbol{q};\Gamma_{\chi})
    \min_{\boldsymbol{y} \in \Gamma_{\chi}}
    \left\|
    \boldsymbol{q} - \boldsymbol{y}
    \right\|_2 .
    \label{eq:sdf_def}
\end{equation}
Here, $\sigma(\boldsymbol{q};\Gamma_{\chi})$ is the sign function defined with respect to the reconstructed surface $\Gamma_{\chi}$,
\begin{equation}
    \sigma(\boldsymbol{q};\Gamma_{\chi})
    =
    \begin{cases}
    +1, & \boldsymbol{q} \ \text{is outside the reconstructed geometry}, \\
    -1, & \boldsymbol{q} \ \text{is inside the reconstructed geometry}.
    \end{cases}
    \label{eq:sign_function}
\end{equation}
Byconstruction, the SDF $\phi$ preserves the interface obtained from the Poisson field:
\begin{equation}
    \left\{
    \boldsymbol{q}
    \; \big| \;
    \phi(\boldsymbol{q}) = 0
    \right\}
    =
    \Gamma_{\chi},
\label{eq:zero_level_set}
\end{equation}
while replacing the implicit indicator values of $\chi_h$ with signed Euclidean distances to the surface.
A signed distance field satisfies the Eikonal property
\begin{equation}
    \left\|
    \nabla \phi(\boldsymbol{q})
    \right\|_2 = 1
    \quad \text{almost everywhere},
    \label{eq:eikonal}
\end{equation}
except near non-smooth surface features and medial-axis locations where the closest surface point is not unique. This property ensures that the gradient $\nabla\phi$ provides a stable unit normal to the surface, making the SDF particularly valuable for physics-based numerical simulation. Specifically, distance queries, surface-normal estimation via $\nabla\phi / \|\nabla\phi\|$, curvature computation via $\text{div}(\nabla\phi)$, and immersed-boundary methods for complex geometries all rely on the Eikonal property for stability and geometric consistency~\citep{peskin2002ibm,mittal2005ibm}.

Computing the SDF directly from the implicit field $\chi_h$ would require solving a point-to-implicit-surface distance problem for every query location, which is computationally expensive and difficult to parallelize. Instead, we leverage the explicit triangular surface $\Gamma_{\chi}$ extracted via Marching Cubes. This mesh representation enables efficient spatial queries: the closest point on the surface can be found via distance-to-triangle computations, which are standard computational geometry operations well-suited to GPU parallelization. Specifically, one GPU thread is assigned to each query point $\boldsymbol{q}$. For each query point, the closest point on the reconstructed triangular surface is obtained by minimizing the distance to all candidate triangles,
\begin{equation}
    d(\boldsymbol{q},\Gamma_{\chi})
    =
    \min_{T_k \in \Gamma_{\chi}}
    d(\boldsymbol{q},T_k),
    \label{eq:mesh_distance}
\end{equation}
where $T_k$ denotes the $k$th triangle extracted from $\chi_h$ and $d(\boldsymbol{q},T_k)$ is the point-to-triangle distance. For a given triangle, the query point is first projected onto the triangle plane. If the projection lies inside the triangle, the point-to-triangle distance is the normal distance to the plane. Otherwise, the distance is computed as the minimum distance to the triangle edges and vertices.

The sign of the distance is determined with respect to the same reconstructed surface $\Gamma_{\chi}$. In this work, we use the parity-based inside--outside test introduced in Section~\ref{subsec:normal_orientation}: a ray is cast from the query point, and the parity of the number of surface intersections determines whether the point is inside or outside the geometry. The final SDF is then given by
\begin{equation}
    \phi(\boldsymbol{q})
    =
    \sigma(\boldsymbol{q};\Gamma_{\chi})
    d(\boldsymbol{q},\Gamma_{\chi}).
    \label{eq:sdf_final}
\end{equation}
The resulting SDF $\phi$ is a volumetric scalar field on the original computational grid, where each grid point contains the signed distance to the Poisson-reconstructed surface. This field preserves the interface geometry defined by $\Gamma_{\chi}$ (its zero level set) while providing the metrically consistent distance information required for downstream immersed-boundary simulations and gradient-based shape optimization. The computation is fully data-parallel over query points and is implemented on GPU.

\subsection{Differentiability}
\label{sec:AD}
The differentiability of \emph{Warp-Geo} is enabled by combining \texttt{Warp} and \texttt{JAX}. Most geometry-processing operations, including normal estimation, SDF evaluation, and other point-wise or grid-based kernels, are implemented in \texttt{Warp} and recorded by \texttt{Warp.Tape}. The Poisson reconstruction step requires solving a large sparse linear system and is therefore handled in \texttt{JAX}, where robust Krylov-subspace and multi-grid iterative linear solvers are available~\cite{LIU2026102966}.

To optimize the reconstructed geometry, we need to compute gradients of a scalar loss function $\mathcal{L}$ with respect to the design parameters $\boldsymbol{\theta}$. Specifically, we consider the parameterized Poisson system
\begin{equation}
    \boldsymbol{A}(\boldsymbol{\theta})
    \boldsymbol{a}(\boldsymbol{\theta})
    =
    \boldsymbol{b}(\boldsymbol{\theta}),
    \label{eq:linear_ad}
\end{equation}
where: $\boldsymbol{A} \in \mathbb{R}^{N_g \times N_g}$ is the stiffness matrix from the Galerkin discretization, $\boldsymbol{a}, \boldsymbol{b} \in \mathbb{R}^{N_g}$ are the unknown coefficients 
and right-hand-side vector, respectively, $\boldsymbol{\theta} \in \mathbb{R}^{N_\theta}$ denotes the design parameters (e.g., point-cloud coordinates or shape-control parameters), $N_g$ is the number of basis-function coefficients,and $N_\theta$ is the number of optimizable geometry parameters.

The gradient of the loss function with respect to the design parameters is computed by the chain rule,
\begin{equation}
    \frac{\partial \mathcal{L}}{\partial \boldsymbol{\theta}}
    =
    \left(\frac{\partial \mathcal{L}}{\partial \boldsymbol{a}}\right)^T
    \frac{\partial \boldsymbol{a}}{\partial \boldsymbol{\theta}},
    \label{eq:chain_rule}
\end{equation}
where $\partial \mathcal{L} / \partial \boldsymbol{a} \in \mathbb{R}^{N_g}$ is computed via automatic differentiation through surface extraction and SDF evaluation, while the full Jacobi matrix ${\partial \boldsymbol{a}} / {\partial \boldsymbol{\theta}} \in \mathbb{R}^{N_g \times N_\theta}$ is very high-dimensional and expensive to compute directly. To avoid this, we use 
implicit differentiation.

Differentiating Eq.~\eqref{eq:linear_ad} with respect to $\boldsymbol{\theta}$ yields
\begin{equation}
    \boldsymbol{A}(\boldsymbol{\theta})
    \frac{\partial \boldsymbol{a}}{\partial \boldsymbol{\theta}}
    +
    \frac{\partial \boldsymbol{A}}{\partial \boldsymbol{\theta}} \boldsymbol{a}
    =
    \frac{\partial \boldsymbol{b}}{\partial \boldsymbol{\theta}}.
    \label{eq:linear_sensitivity}
\end{equation}
Solving this for all $N_\theta$ parameters is prohibitively expensive. Instead, we 
employ the adjoint method. Introduce the adjoint variable $\boldsymbol{\lambda}$ 
satisfying
\begin{equation}
    \boldsymbol{A}(\boldsymbol{\theta})^T \boldsymbol{\lambda}
    =
    \left(\frac{\partial \mathcal{L}}{\partial \boldsymbol{a}}\right)^T,
    \label{eq:adjoint_linear}
\end{equation}
Multiplying Eq.~\eqref{eq:linear_sensitivity} by $\boldsymbol{\lambda}^T$ 
and using Eq.~\eqref{eq:adjoint_linear}, the gradient becomes
\begin{equation}
    \frac{\partial \mathcal{L}}{\partial \boldsymbol{\theta}}
    =
    \boldsymbol{\lambda}^T
    \left(
    \frac{\partial \boldsymbol{b}}{\partial \boldsymbol{\theta}}
    -
    \frac{\partial \boldsymbol{A}}{\partial \boldsymbol{\theta}} \boldsymbol{a}
    \right).
    \label{eq:linear_ad_gradient}
\end{equation}
This requires only one additional linear solve, independent of $N_\theta$.

In practice, \texttt{Warp.Tape} records the differentiable operations before and after the Poisson solve. During the forward pass, \emph{Warp-Geo} assembles $\boldsymbol{A}$ and $\boldsymbol{b}$, solves Eq.~\eqref{eq:linear_ad} in \texttt{JAX}, and then proceeds with surface extraction and SDF computation. During the backward pass, Eq.~\eqref{eq:adjoint_linear} is solved as an additional linear system, and Eq.~\eqref{eq:linear_ad_gradient} propagates gradients through the Poisson reconstruction. This treatment avoids storing the full solver trajectory and enables \emph{Warp-Geo} to be coupled with downstream differentiable computational solvers for end-to-end geometry optimization.

\textit{Warp-Geo} is differentiable with respect to the input point-cloud coordinates. Nevertheless, directly treating all point coordinates as optimization variables results in a high-dimensional design space that can be difficult to regularize. For practical inverse-design problems, it is more effective to parameterize the point cloud or the underlying geometry using a compact set of design variables in \texttt{Warp} or \texttt{JAX}. Gradients can then be propagated through the full reconstruction pipeline to these parameters, leading to a more stable and physically meaningful optimization formulation. Beyond parameter gradients, \emph{Warp-Geo} also supports differentiation of the SDF with respect to spatial query points, similar to existing differentiable SDF query tools such as \texttt{pytorch-volumetric}. This combination of geometry-parameter differentiation and query-point differentiation is one of the key capabilities of the proposed framework.

\section{Results and discussion}
\label{sec:results}

\subsection{Forward shape representation}
\label{sec:forward}

In this section, we reconstruct SDFs from point clouds of representative geometries using \emph{Warp-Geo} and evaluate key properties of the reconstructed fields.

\begin{figure}[htp!]
    \centering
    \includegraphics[width=1.0\textwidth]{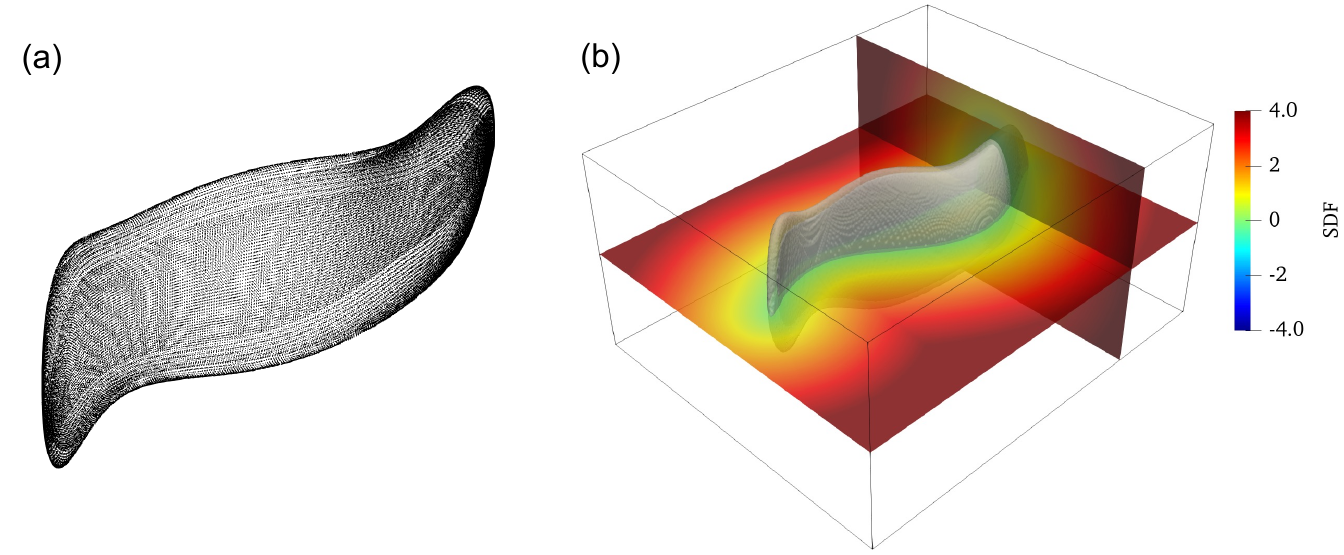}
    \caption{SDF reconstruction for the fish geometry.
    (a) Input point cloud. 
    (b) Reconstructed SDF with the zero-level isosurface extracted at $\phi=0$.}
    \label{fig:fish_sdf_gen}
\end{figure}
Figure~\ref{fig:fish_sdf_gen} shows the input point cloud of a thin fish body and the corresponding reconstructed surface. The reconstruction uses a grid of $338 \times 120 \times 151$ cells, controlled by hyperparameter $d$ in Table~\ref{tab:hyperparameters}, with the computational domain automatically generated from the point-cloud bounding box or manually specified according to the target simulation domain. As shown in Fig.~\ref{fig:fish_sdf_gen}(b), the reconstructed surface captures main geometric features, including the thin body and sharp fins.

\begin{figure}[htp!]
    \centering
    \includegraphics[width=1.0\textwidth]{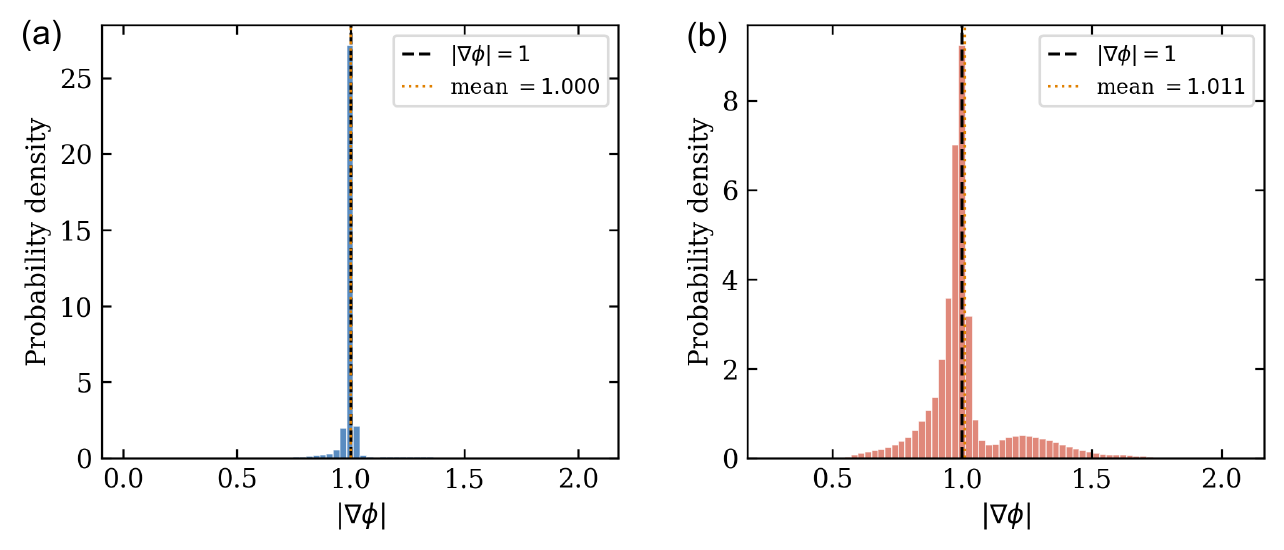}
    \caption{Properties of the reconstructed SDF for the fish geometry. 
    (a) Probability density function (PDF) of $|\nabla \phi|$ over the entire computational domain. 
    (b) PDF of $|\nabla \phi|$ near the geometry surface.}
    \label{fig:fish_sdf_property}
\end{figure}

The quality of the reconstructed SDF is evaluated in Fig.~\ref{fig:fish_sdf_property}. 
Figure~\ref{fig:fish_sdf_property}(a) shows the probability density function (PDF) of $|\nabla \phi|$ over the entire computational domain. The distribution is concentrated around one, confirming that the reconstructed field satisfies the Eikonal property globally. To examine the near-surface region more closely, Fig.~\ref{fig:fish_sdf_property}(b) reports the PDF of $|\nabla \phi|$ near the geometry surface. The mean remains near one, although a small tail appears in the distribution. This deviation occurs in regions with sharp features, such as the thin fish tail, where closest-point projection becomes non-unique and the SDF gradient degrades locally, as discussed in Section~\ref{subsec:surface_sdf}.

To further assess the robustness and generality of \emph{Warp-Geo} across diverse geometries, we test it on five additional representative geometries: a bio-inspired shape, an airplane, an automobile, a jet aircraft, and a vascular geometry. These cases span the geometric complexities encountered in CFD applications: non-convex surfaces, thin structures such as aircraft wings and fins, and sharp features.

\begin{figure}[htp!]
    \centering
    \includegraphics[width=1.0\textwidth]{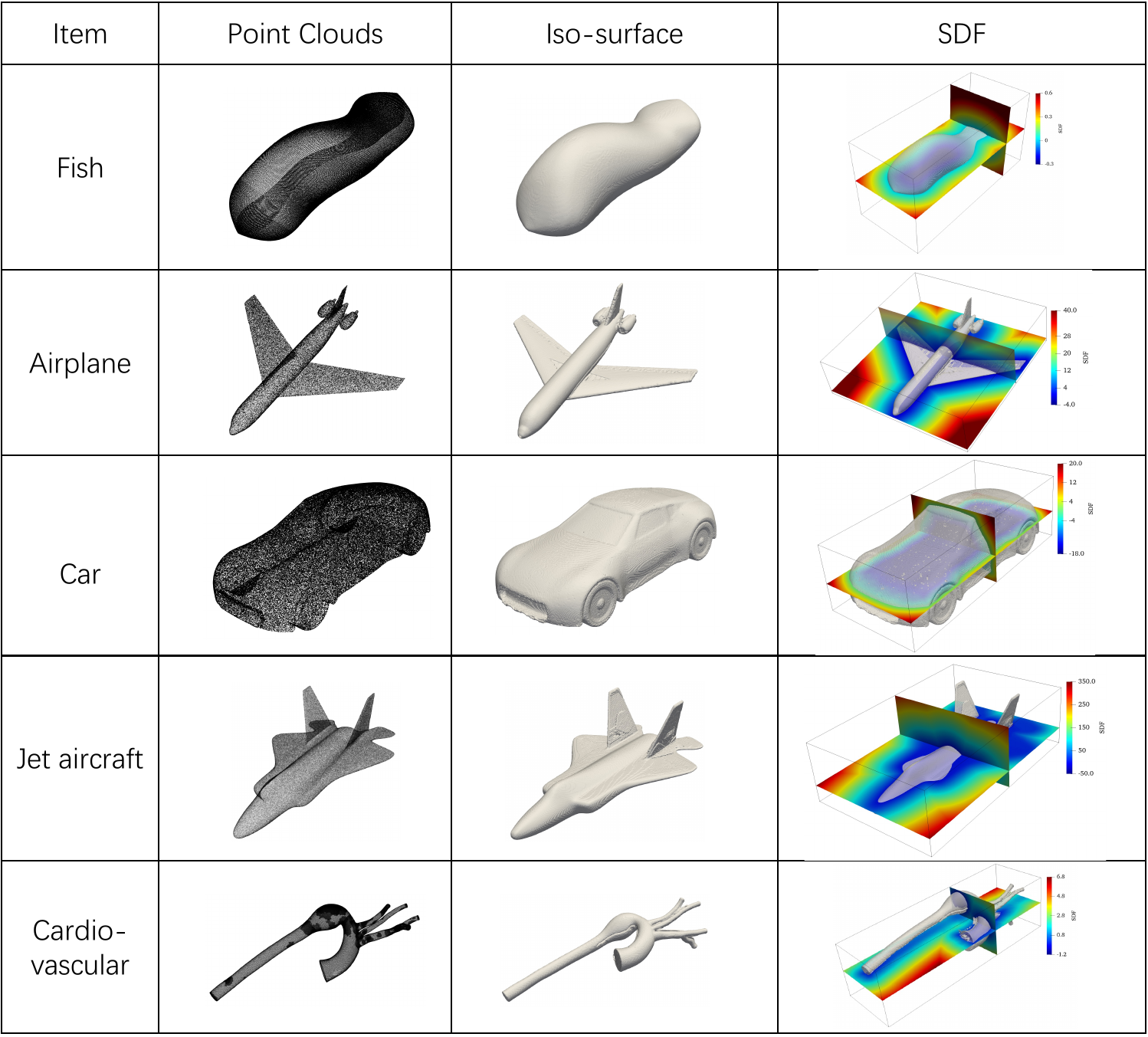}
    \caption{Reconstruction results for multiple point-cloud geometries using \emph{Warp-Geo} framework. The first row shows the input point clouds, the second row shows the extracted iso-surfaces, and the third row shows representative slices of the reconstructed SDFs. The tested cases include bio-inspired, aircraft, automobile, jet, and cardiovascular geometries with different levels of geometric complexity.}
    \label{fig:all_sdf}
\end{figure}
As shown in Fig.~\ref{fig:all_sdf}, \emph{Warp-Geo} successfully reconstructs both the surface and the corresponding SDF from the input point clouds for all tested geometries. The reconstructed isosurfaces preserve the main geometric features, and the computed SDFs provide smooth distance variations around the objects. Some localized non-smoothness can still be observed in regions with very thin or closely spaced structures. This is expected because nearby surface sheets may become difficult to distinguish during normal-vector estimation and normal orientation, especially when the local point-cloud spacing is comparable to the thickness of the geometric feature. Nevertheless, the results demonstrate that \emph{Warp-Geo} can robustly handle diverse point-cloud geometries and generate SDF representations suitable for downstream immersed-boundary and inverse-design applications. Additional examples are provided in our open-source codebase.

\subsection{Dynamic-boundary simulation: integration with a CFD solver}
\label{sec:embedding}

We couple \emph{Warp-Geo} with a downstream differentiable CFD solver~\citep{fan2026diff} to demonstrate its effectiveness in dynamic-boundary simulation. The CFD solver details are in~\citep{fan2026diff}. The prescribed fish trajectory and body deformation follow prior studies~\citep{fishshape,kern2006simulations}, with geometry imposed as a moving no-slip boundary via a sharp immersed-boundary method.

\begin{figure}[htp!]
    \centering
    \includegraphics[width=0.9\textwidth]{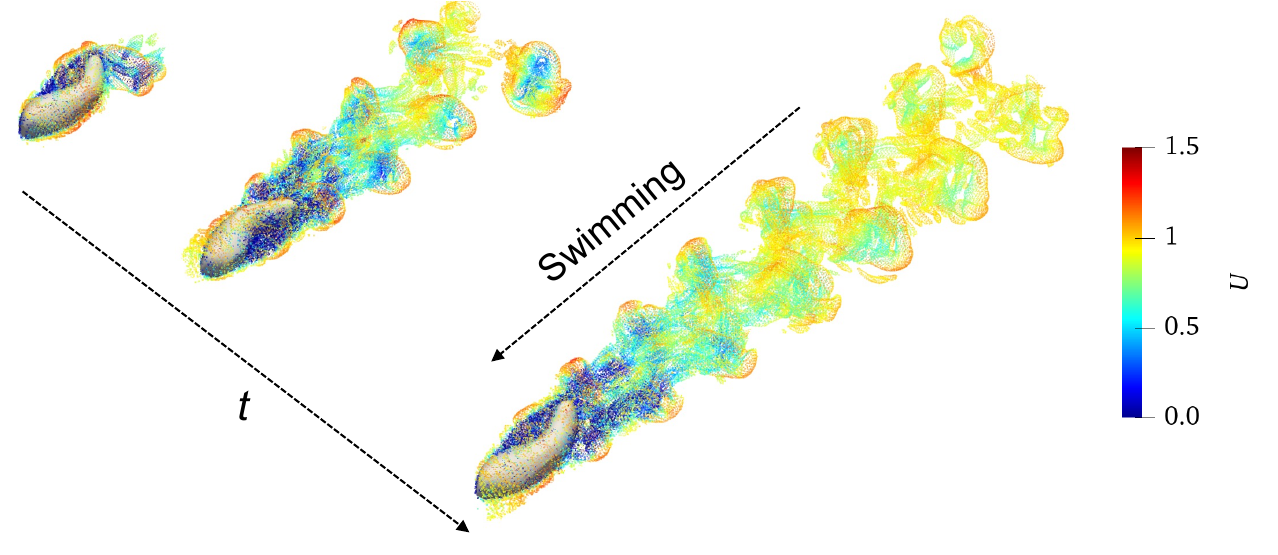}
    \caption{Coupling \emph{Warp-Geo} with a downstream differentiable CFD solver for swimming-fish simulation. The SDF is dynamically reconstructed from the prescribed fish motion and used as the moving immersed boundary.}
    \label{fig:fish_cfd}
\end{figure}

Conventional level-set and SDF representations update the geometry field for moving boundaries by solving advection equations followed by reinitialization~\citep{cui2018sharp}. In contrast, \emph{Warp-Geo} efficiently regenerates the SDF directly from the updated point-cloud representation at each time step, eliminating the need for advection and 
reinit operations. Figure~\ref{fig:fish_cfd} shows the vortex wake generated by the swimming fish, confirming that the reconstructed SDF provides a valid and effective geometry representation for CFD simulation. This example demonstrates the potential of \emph{Warp-Geo} for future end-to-end differentiable workflows, where geometry representation, 
flow simulation, and shape optimization can be tightly coupled in a unified framework.

\subsection{Gradient-based shape optimization}
\label{sec:inverse}

We validate \emph{Warp-Geo}'s differentiability through an inverse geometry optimization task. In a full design workflow, \emph{Warp-Geo} couples with a downstream differentiable solver to optimize physical quantities such as drag reduction or aerodynamic performance. Here, we focus on the geometry reconstruction and assume gradients propagate to the SDF. The inverse problem is thus simplified to morphing an initial geometry toward a target by matching their SDFs.

Directly optimizing all point-cloud coordinates is possible in our framework, but it creates a high-dimensional, weakly constrained design space. Instead, we parameterize the fish geometry using a compact set of shape parameters. Following prior geometric descriptions~\citep{carling1998self,kern2006simulations}, the fish point cloud is controlled by six width and eight height parameters for prescribed body length. This low-dimensional parameterization provides stable optimization while allowing effective body deformation.

\begin{figure}[htp!]
    \centering
    \includegraphics[width=1.0\textwidth]{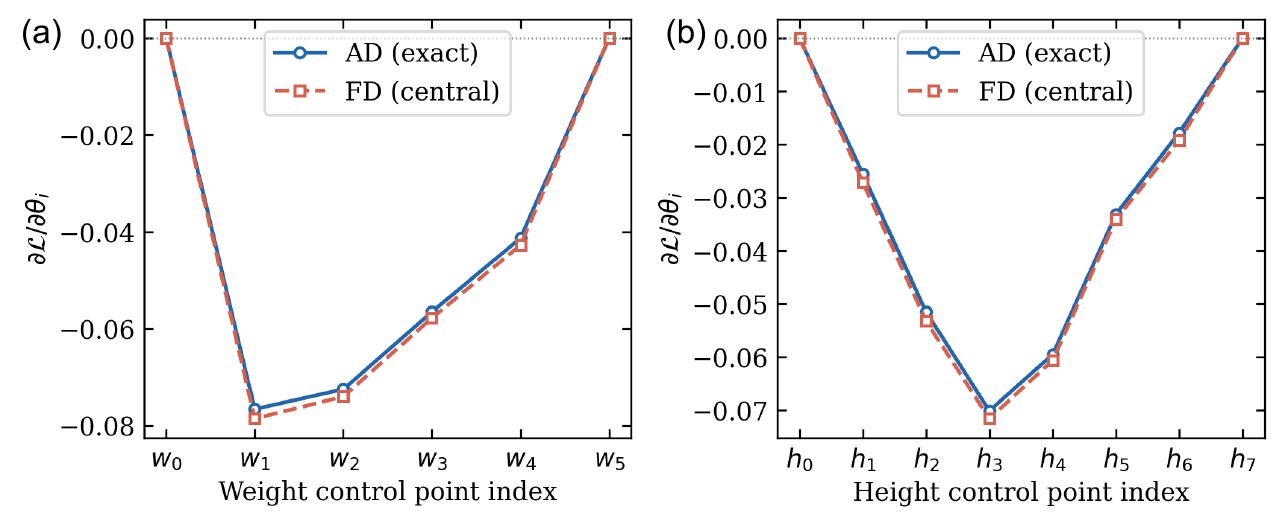}
    \caption{Comparison of gradients from automatic differentiation (AD) and finite difference (FD): 
    (a) with respect to width control point $w_i$; 
    (b) with respect to height control point $h_i$.}
    \label{fig:AD_verify}
\end{figure}

We verify the AD gradients by comparing them with finite-difference estimates for width and height parameters. The loss function matches the reconstructed SDF to a target SDF:
\begin{equation} 
    \mathcal{L}(w_i, h_i) = \left\| \tilde{\phi} - \phi(w_i, h_i) \right\|_2^2,
    \label{eq:sdf_matching_loss} 
\end{equation} 
where $\tilde{\phi}$ is the target SDF and $\phi(w_i, h_i)$ is the reconstructed SDF parameterized by width $w_i$ and height $h_i$ parameters.
Figure~\ref{fig:AD_verify} shows excellent agreement between AD and finite-difference gradients, confirming the correctness of the differentiable reconstruction pipeline. Importantly, AD evaluation is substantially more efficient, avoiding repeated forward evaluations per parameter. In this example, AD computes the full gradient approximately 15 times faster than finite difference.

\begin{figure}[H]
    \centering
    \includegraphics[width=1.0\textwidth]{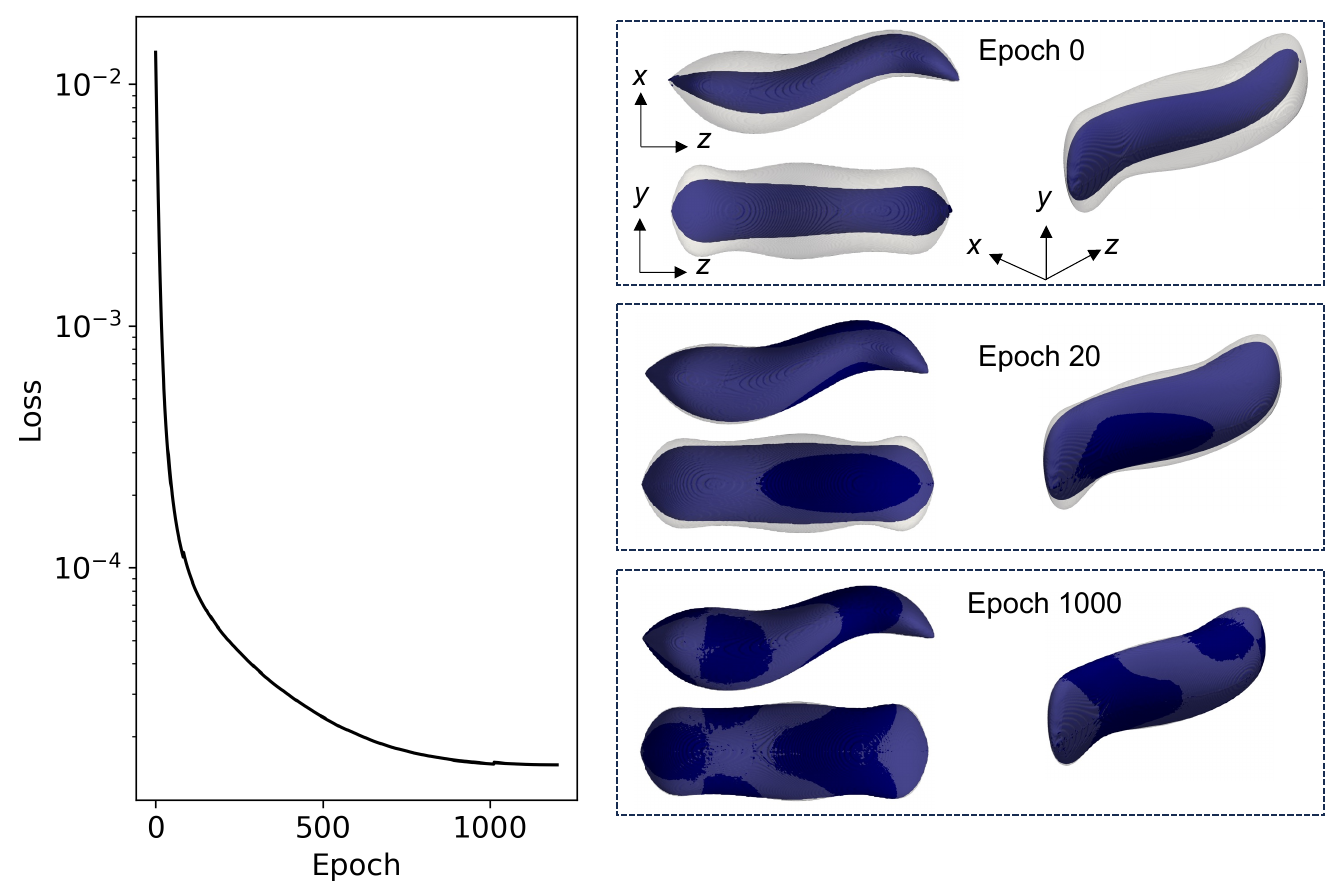}
    \caption{Optimization history shown as 3D views at selected epochs. Starting from an inaccurate initialization (Epoch 0), the geometry converges toward the target using automatic-differentiation gradients. Dark blue denotes the optimized shape; light gray denotes the target.}
    \label{fig:shape_optimize}
\end{figure}

\begin{figure}[H]
    \centering
    \includegraphics[width=0.9\textwidth]{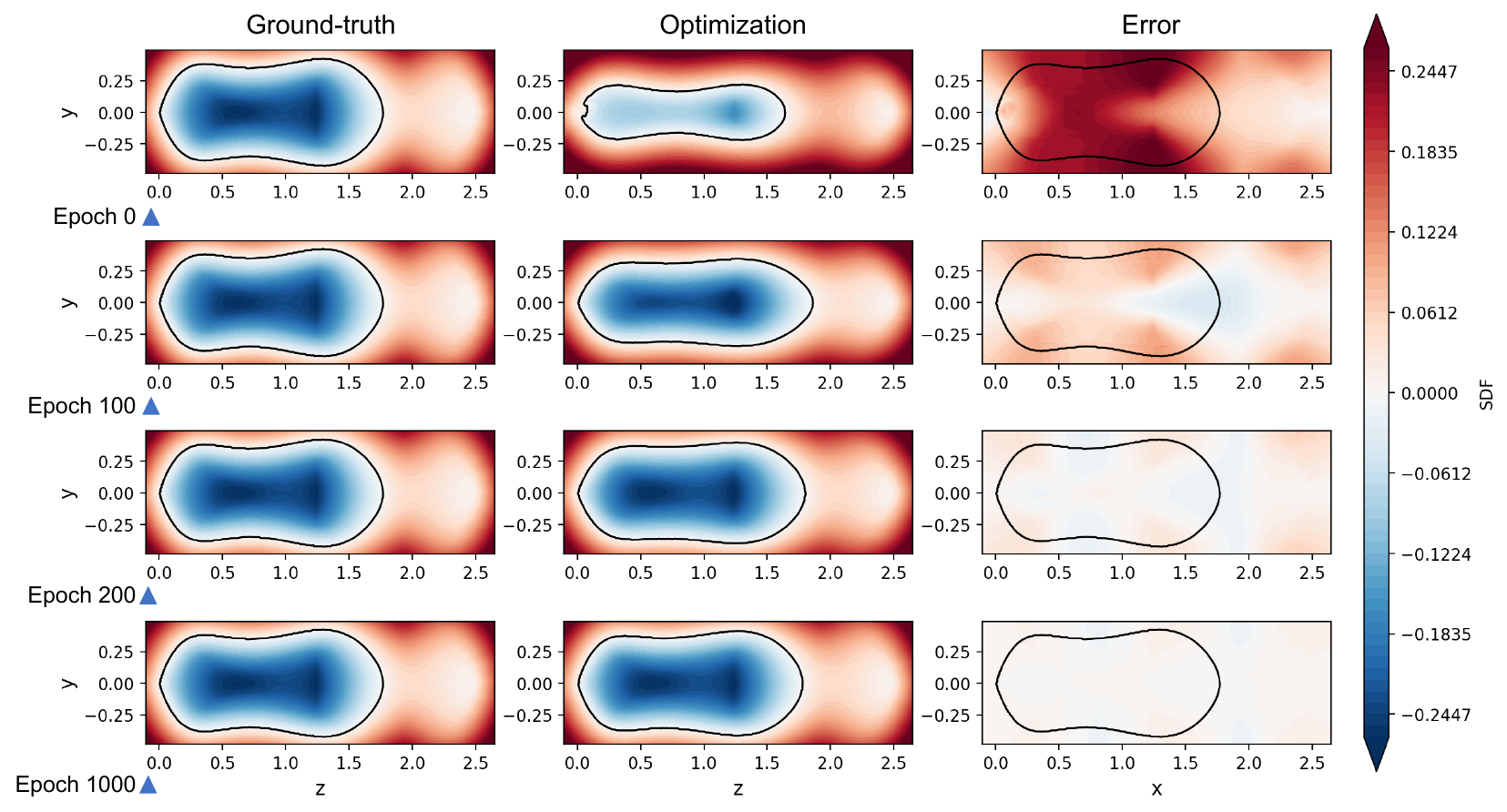}
    \caption{Optimization history shown as SDF cross-sections in the $y$--$z$ plane. The black contour marks the zero isosurface $\phi = 0$.}
    \label{fig:sdf_optimize}
\end{figure}

\begin{figure}[H]
    \centering
    \includegraphics[width=0.9\textwidth]{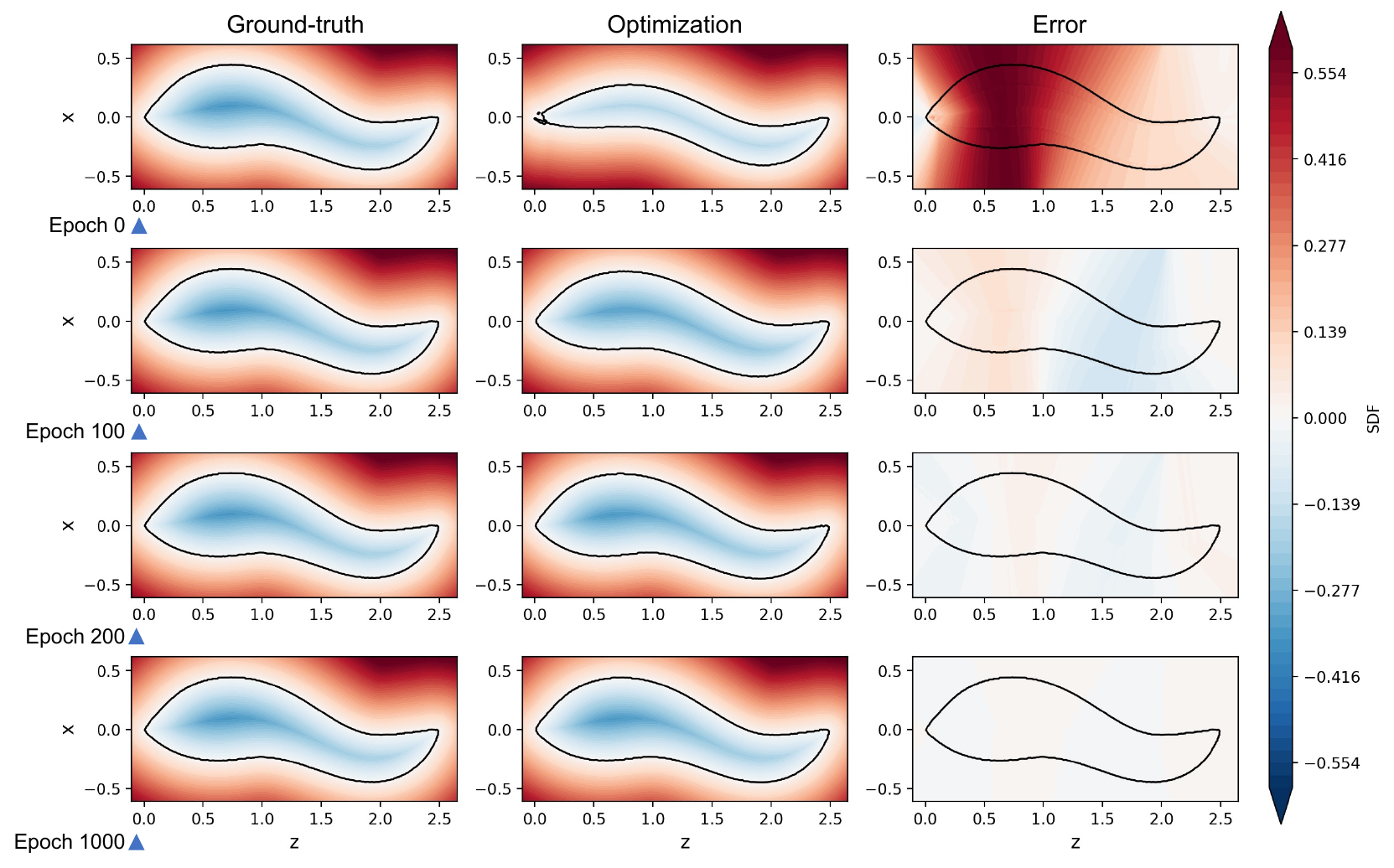}
    \caption{Optimization history shown as SDF cross-sections in the $x$--$z$ plane. The black contour marks the zero isosurface $\phi = 0$.}
    \label{fig:sdf_optimize_view2}
\end{figure}

We then optimize width and height parameters from random initialization. Figure~\ref{fig:shape_optimize} shows the optimization history. After 20 epochs, the optimized geometry captures the overall fish profile, though local features remain under-resolved. After 500 epochs, both global shape and local details are accurately recovered. The corresponding SDF slices and error fields (Figs.~\ref{fig:sdf_optimize} and~\ref{fig:sdf_optimize_view2}) confirm convergence toward the target. These results demonstrate that \emph{Warp-Geo} provides reliable gradients for parameterized geometries and enables inverse shape optimization through its differentiable SDF representation.

\subsection{Cost evaluation}
\label{sec:cost}

We evaluate the computational cost and scalability of \emph{Warp-Geo} using the fish geometry. The accuracy of implicit surface representation depends on background grid resolution, controlled by hyperparameter $d$, which specifies grid density along each coordinate direction and determines the total number of grid points.

\begin{figure}[htp!]
    \centering
    \includegraphics[width=1.0\textwidth]{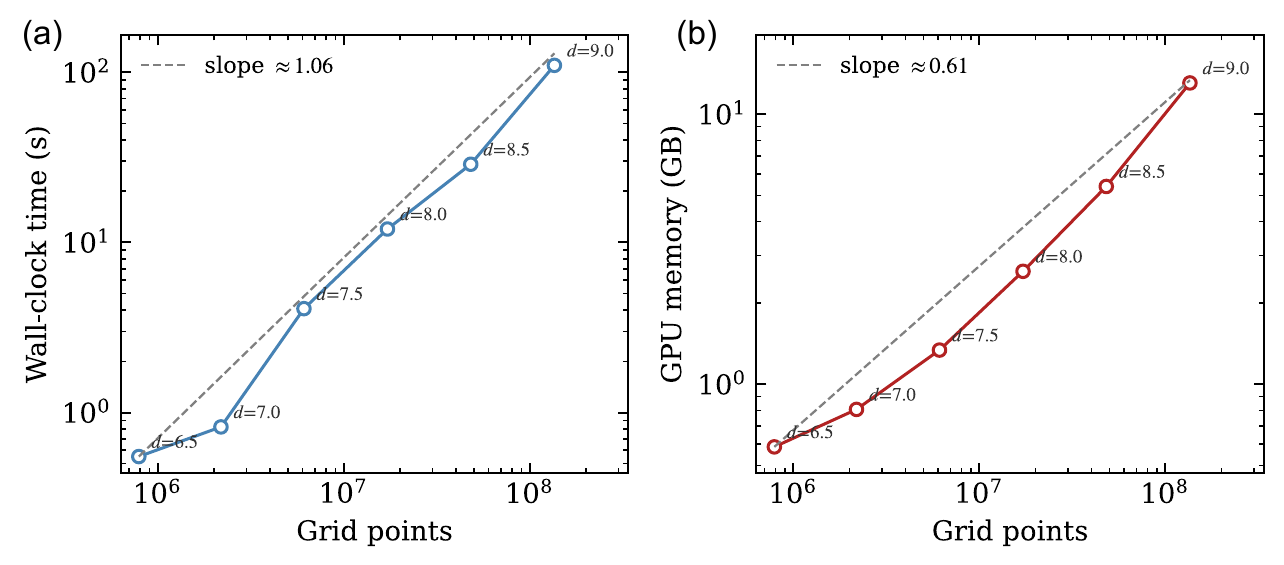}
    \caption{Computational cost of SDF reconstruction (fish point cloud) at different grid resolutions: 
    (a) wall-clock time versus total number of grid points; 
    (b) GPU memory versus total number of grid points.}
    \label{fig:cost}
\end{figure}

Figure~\ref{fig:cost} shows wall-clock time and GPU memory usage for different values of $d$. Figure~\ref{fig:cost}(a) shows that wall-clock time scales approximately linearly with the total number of grid points on a log--log plot, with fitted slope close to one. Figure~\ref{fig:cost}(b) shows GPU memory exhibits similar scaling, though slightly sublinear over the tested range. These results indicate favorable scalability with respect to grid resolution. The overall computational cost remains practical, making \emph{Warp-Geo} a suitable differentiable geometry module for coupling with downstream computational solvers.

\section{Conclusions}
\label{sec:conclusion}

We present \emph{Warp-Geo}, a differentiable GPU-accelerated framework for reconstructing signed distance fields with surface normal from point clouds. The framework integrates normal estimation, orientation correction, Poisson surface reconstruction, isosurface extraction, and SDF reinitialization into a unified pipeline. A uniform-grid discretization with dense 27-cell normal distribution overcomes weak coupling inherent in naive uniform grids, achieving robust surface reconstruction while maintaining GPU parallelization efficiency.

Comprehensive validation across diverse geometries—fish, aircraft, automobile, jet, and cardiovascular cases—demonstrates accurate SDF reconstruction. The reconstructed fields satisfy the Eikonal property globally, with localized degradation only at sharp features is expected. Computational cost scales favorably with grid resolution, confirming practical scalability. The framework enables dynamic-boundary CFD simulation through direct SDF regeneration at each time step, eliminating advection and reinitialization overhead compared to conventional level-set methods. Inverse shape-optimization experiments validate end-to-end differentiability: automatic-differentiation gradients agree with finite-difference estimates while being approximately 15 times faster.

\emph{Warp-Geo} provides a practical differentiable geometry representation that seamlessly integrates with downstream differentiable solvers, enabling end-to-end shape optimization and physics-driven design workflows. The framework addresses a critical gap in differentiable computational modeling by offering robust, efficient, and flexible geometry handling. Future directions include adaptive resolution for multiscale features, handling topology changes through levelset dynamics, and integration with differentiable physics solvers for 
coupled aerodynamic and structural optimization.

\section*{Acknowledgements}
The authors would like to acknowledge the funds from the Department of Energy, as part Genesis Mission project (Award No. DE-SC0026662) and ASCR project (Award No. DE-SC0026687), as well as the funds from the National Science Foundation (Award No. OAC-2047127). This research is supported by the NVIDIA Academic Grant Program using NVIDIA RTX PRO 6000 GPUs, NVIDIA \texttt{CUDA} and \texttt{Warp} Toolkits.
 
\section*{Compliance with Ethical Standards}
Conflict of Interest: The authors declare that they have no conflict of interest.

\bibliographystyle{elsarticle-num}
\bibliography{ref,own-ref}

\end{document}